\documentclass[%
superscriptaddress,
preprint,
 amsmath,amssymb,
 aps,
pra,
]{revtex4-2}

\usepackage{graphicx}
\usepackage{dcolumn}
\usepackage{bm}
\usepackage{braket}
\usepackage{comment} 
\usepackage{hyperref}
\usepackage{siunitx}

\usepackage{xcolor}
\newcommand{\red}[1]{\textcolor{black}{#1}}

\usepackage{nicematrix}
\NiceMatrixOptions{cell-space-limits = 1pt}

\usepackage{mathtools}

\usepackage{amsthm}

\newtheorem*{theorem*}{Theorem}

\newtheorem{algorithm}{Algorithm}

\theoremstyle{definition}

\theoremstyle{remark}

\newtheorem*{example*}{Example}
\newtheorem*{algorithm*}{Algorithms}

\usepackage{etoolbox}

\makeatletter
\patchcmd{\frontmatter@abstract@produce}
  {\vskip200\p@\@plus1fil
   \penalty-200\relax
   \vskip-200\p@\@plus-1fil}
  {}
  {}
  {}
\makeatother

\begin{document}

\title{Unitary control of partially coherent waves. I. Absorption}

\author{Cheng Guo}
\email{guocheng@stanford.edu}
\affiliation{Ginzton Laboratory and Department of Electrical Engineering, Stanford University, Stanford, California 94305, USA}

\author{Shanhui Fan}
\email{shanhui@stanford.edu}
\affiliation{Ginzton Laboratory and Department of Electrical Engineering, Stanford University, Stanford, California 94305, USA}%

\date{\today}

\begin{abstract}
The coherent control of wave absorption has important applications in areas such as energy harvesting, imaging, and sensing. However, most practical scenarios involve the absorption of partially coherent rather than fully coherent waves. Here we present a systematic theory of unitary control over the absorption of partially coherent waves by linear systems. Given an absorbing system and incident partially coherent wave, we provide analytical expressions for the range of attainable absorptivity under arbitrary unitary transformations of the incident field. We also present an explicit algorithm to construct the unitary control scheme that achieves any desired absorptivity within that attainable range. As applications of our theory, we derive the conditions required for achieving two new phenomena - partially coherent perfect absorption and partially coherent zero absorption. Furthermore, we prove a theorem relating the coherence properties of the incident field, as quantified by majorization, to the resulting absorption intervals. Our results provide both fundamental insights and practical prescriptions for exploiting unitary control to shape the absorption of partially coherent waves. The theory applies across the electromagnetic spectrum as well as to other classical wave systems such as acoustic waves.
\end{abstract}
\maketitle


\section{introduction}\label{sec:introduction}

Absorption is a fundamental wave phenomenon~\cite{planck1991,chen2005,zhang2007,howell2016,fan2017,cuevas2018b,li2021e}. Controlling wave absorption has significant implications in diverse applications, including renewable energy~\cite{ottens2011, guha2012, babuty2013, rephaeli2013, raman2014, boriskina2016,raj2017a,fiorino2018a,park2021,zhu2019c,li2019g}, imaging~\cite{kittel2005}, and sensing~\cite{muraviev2018,tan2020}.

Approaches to controlling wave absorption can be classified into two categories: structural design and wave control. In the structural design approach, desired absorption behaviors are achieved by designing the absorber's structure. For instance, recent progress in thermal photonics has enabled novel photonic absorbers that differ drastically from traditional blackbody absorbers~\cite{greffet2002a, guo2012b, dezoysa2012, yu2013, pelton2015,  thompson2018, baranov2019a, guo2021a}. Photonic structures can be designed with narrowband absorptivity in either the spectral or angular domain~\cite{greffet2002a}, or with low absorptivity at solar wavelengths and high emissivity in the mid-infrared for daytime radiative cooling~\cite{rephaeli2013,raman2014}.

In the wave control approach, desired absorption behaviors are achieved by manipulating the external waves interacting with the absorber. This approach is termed unitary control, as it involves a unitary transformation of the external waves~\cite{guo2023b}. Unitary control is widely used in applications where the absorber structure is fixed. For example, Fresnel lenses are used as concentrators to enhance solar cell absorption by reshaping the external waves~\cite{green2006}. Unitary control has also been used to achieve novel absorption phenomena like coherent perfect absorption~\cite{chong2010a,wan2011,sun2014,baranov2017,mullers2018a,pichler2019a,sweeney2019a,chen2020ac,wang2021h,slobodkin2022}.

A systematic theory of unitary control has recently been developed for coherent waves~\cite{guo2023b}. However, many practical applications, such as solar energy and radio astronomy, involve the absorption of partially coherent waves~\cite{mandel1995,goodman2000}, as many wave sources are inherently partially coherent. In order to develope a theory of unitary control for manipulation of the absorption of partially coherent wave, one needs to take into account the interplay between the properties of the absorbers, and the coherence property of the incident waves. Such a theory thus represents a step forward beyond the theory of unitary control for coherent waves.

In this paper, we develop a systematic theory of unitary control for partially coherent waves. Our theory addresses two basic questions: (i) Given an absorbing object and an incident partially coherent wave, what is the range of all attainable absorptivity under unitary control? (ii) How can a given absorptivity be obtained via unitary control? The first question addresses the capability and limitations of unitary control in absorptivity, while the second focuses on implementation. In this paper, we provide complete answers to both questions. As physical applications of our theory, we determine the conditions for two new phenomena: partially coherent perfect absorption and partially coherent zero absorption. We also examine how the degree of coherence, measured by the majorization order, affects the attainable absorptivity and prove that majorized coherence implies nested absorption intervals.

This paper is the first in a series on the unitary control of partially coherent waves. In this paper, we investigate the unitary control of the absorption of partially coherent waves. In the second paper~\cite{guo2024b}, we further extend the unitary control method to manipulate the transmission and reflection of partially coherent waves. We have adopted the same mathematical notations and similar proof techniques throughout this series of papers. 

The rest of this paper is organized as follows. In Sec.~\ref{sec:notations}, we summarize useful mathematical notations. In Sec.~\ref{sec:theory}, we develop a general theory of unitary control over partially coherent wave absorption. In Sec.~\ref{sec:applications}, we discuss the physical applications of our theory. We conclude in Sec.~\ref{sec:conclusion}. 

\section{Notations}\label{sec:notations}
We first summarize the notations related to matrices. We denote by $M_{n}$ the set of $n\times n$ complex matrices and  $U(n)$ the set of $n\times n$ unitary matrices. For $M \in M_{n}$, we denote by $\bm{d} (M) = \left(d_{1}(M),\ldots, d_{n}({M})\right)^{T}$, $\bm{\lambda} (M) = \left(\lambda_{1}(M), \ldots, \lambda_{n}(M)\right)^{T}$, and $\bm{\sigma} (M) = \left(\sigma_{1}(M),\ldots,\sigma_{n}(M)\right)^{T}$ 
the vectors of diagonal entries, eigenvalues, and singular values of $M$~\cite{guo2023c}. We also define the vector 
\begin{equation}
\bm{1} - \bm{\sigma}^2(M) \equiv \left(1 - \sigma^2_{1}(M),\ldots, 1 - \sigma^2_{n}(M)\right).
\end{equation} 

We adopt the following notations for vectors. For $\bm{x} = (x_1, \ldots,x_n)\in \mathbb{R}^n$, we define 
$\bm{x}^\downarrow = (x^\downarrow_1, \ldots,x^\downarrow_n)$ and $\bm{x}^\uparrow = (x^\uparrow_1,\ldots,x^\uparrow_n)$, where $x_1^\downarrow \ge \cdots \ge x_n^\downarrow$ and $x_1^\uparrow \le \cdots \le x_n^\uparrow$ denote the components reordered in non-increasing and non-decreasing orders, respectively. We denote the set of $n$-dimensional probability vectors with components reordered non-increasingly as 
\begin{equation}
\Delta_{n}^\downarrow = \set{ \bm{x} \in \mathbb{R}^{n} | x_{i}\geq 0, x_{i} \geq x_{i+1}, \sum_{i=1}^{n} x_{i} = 1 }.
\end{equation}

\section{Theory}\label{sec:theory}

\subsection{Partially coherent waves}\label{subsec:partially_coherent_waves}

Consider an $n$-dimensional Hilbert space of waves equipped with a set of orthonormal bases. A coherent wave is represented by a complex vector 
\begin{equation}
    \bm{a} = (a_{1}, \dots, a_{n})^T
\end{equation}
A partially coherent wave comprises a statistical mixture of coherent waves and is represented by a density matrix~\cite{landau1981,oneill2003,wolf2003,delimabernardo2017,zhang2019m,korotkova2022} (also known as a coherence~\cite{wolf1985} or coherency~\cite{goodman2000,yamazoe2012,okoro2017} matrix in optics), 
\begin{equation}
    \rho = \langle \bm{a} \bm{a}^{\dagger}\rangle,
\end{equation}
where $\langle \cdot \rangle$ denotes the ensemble average. The density matrix $\rho$ is positive semidefinite. The trace of $\rho$ corresponds to the total power; without loss of generality, we assume it is normalized:
\begin{equation}\label{eq:rho_normalization}
    \operatorname{tr} \rho = 1.
\end{equation}  
The diagonal elements of $\rho$, denoted by
\begin{equation}
\bm{d}(\rho) = (\rho_1, \ldots, \rho_n), 
\end{equation}
represent the power distribution in each basis mode.

The coherence properties are encoded in the eigenvalues of $\rho$, termed the coherence spectrum:
\begin{equation}
\bm{\lambda}^{\downarrow}(\rho) =  (\lambda^{\downarrow}_{1}(\rho),\dots,\lambda^{\downarrow}_{n}(\rho)).   
\end{equation}
$\bm{\lambda}^\downarrow(\rho)$ represents the probability distribution associated with the statistical mixture. A coherent wave has 
\begin{equation}
\bm{\lambda}^\downarrow(\rho) = (1,0,\dots,0),    
\end{equation}
while an incoherent wave has 
\begin{equation}
\bm{\lambda}^\downarrow(\rho) =  (\frac{1}{n}, \frac{1}{n}, \dots, \frac{1}{n}).     
\end{equation}
All waves, including coherent, incoherent, and partially coherent waves, have $\bm{\lambda}^\downarrow(\rho)$ belonging to $\Delta_{n}^\downarrow$.

\begin{figure}[hbtp]
    \centering
    \includegraphics[width=0.5\textwidth]{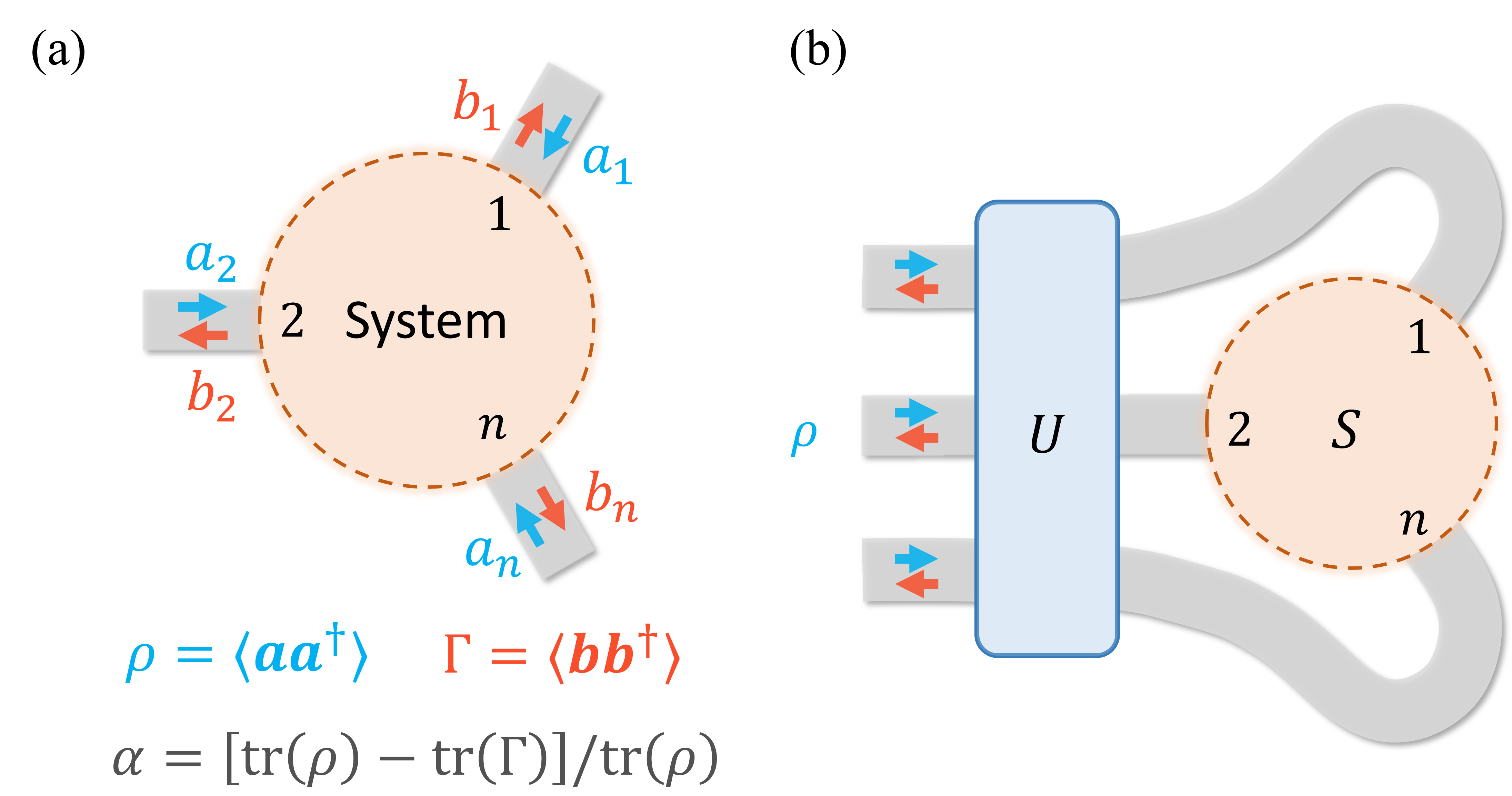}
    \caption{(a) Schematic of partially coherent wave absorption. An input wave characterized by a density matrix $\rho$ is scattered by a passive linear time-invariant system with a scattering matrix $S$, resulting in an output wave with an unnormalized density matrix $\Gamma = S \rho S^\dagger$. The absorptivity $\alpha$ is defined as the ratio between the absorbed power and the input power. (b) Schematic of unitary control of partially coherent wave absorption. A unitary converter $U$ is applied to the input wave before it interacts with the system, enabling the manipulation of the absorptivity $\alpha$. }
    \label{fig:geometry}
\end{figure}

\subsection{Partially coherent wave absorption}

We study the absorption of a partially coherent wave. As shown in Fig.~\ref{fig:geometry}a, we input a partially coherent input wave with an $n \times n$ density matrix $\rho$ into an $n$-port passive linear time-invariant system characterized by a scattering matrix $S \in M_n$~\cite{haus1984}. The output wave is represented by an (unnormalized) density matrix
\begin{equation}\label{eq:Gamma}
    \Gamma = \langle \bm{b} \bm{b}^{\dagger}\rangle = S \rho S^\dagger.
\end{equation}
The trace of $\Gamma$ corresponds to the total output power. The absorptivity $\alpha$ is defined as the ratio between the absorbed power and the input power:
\begin{equation}
    \alpha \coloneqq (\operatorname{tr} \rho - \operatorname{tr} \Gamma) / \operatorname{tr} \rho
\end{equation}
Using Eqs.~(\ref{eq:rho_normalization}) and (\ref{eq:Gamma}), we obtain
\begin{equation}\label{eq:alpha_trace}
\alpha = 1 - \operatorname{tr} \Gamma = \operatorname{tr} (\rho A), 
\end{equation}
with the absorptivity matrix
\begin{equation}
A \coloneqq I -S^{\dagger} S.
\end{equation}
$A$ is positive semidefinite with the \emph{absorption eigenvalues}~\cite{yamilov2016}: 
\begin{equation}\label{eq:lamda_A_sigma_S}
\bm{\lambda}^\downarrow(A) = \bm{1} - \bm{\sigma}^{2\uparrow}(S).  
\end{equation}

\subsection{Unitary control of partially coherent wave absorption}

Now we introduce the concept of unitary control. Unitary control refers to unitarily transforming the input waves using a unitary converter, such as spatial light modulators~\cite{vellekoop2007,popoff2014,yu2017e}, Mach-Zehnder interferometers~\cite{reck1994,miller2013c,miller2013a,miller2013b,carolan2015,miller2015,clements2016,ribeiro2016,wilkes2016,annoni2017,miller2017d,perez2017,harris2018,pai2019}, and multiplane light conversion systems~\cite{morizur2010,labroille2014,tanomura2022,kupianskyi2023,taguchi2023,zhang2023b}. Under unitary control, the input wave is modified via unitary similarity~\cite{horn2012}: 
\begin{equation}
    \rho \to \rho[U] = U \rho U^\dagger.
\end{equation}
Unitary control preserves both the total power and the coherence spectrum of the input wave:
\begin{equation}
\operatorname{tr}\rho[U] = 1, \qquad \bm{\lambda}^\downarrow(\rho[U])=\bm{\lambda}^\downarrow(\rho).   
\end{equation}
Conversely, any pair of waves with the same total power and coherence spectrum can be transformed into each other via unitary control. \red{This follows from the mathematical fact that two Hermitian matrices are unitarily similar if and only if they have the same eigenvalues (see Ref.~\cite{horn2012}, p.~134).} Hence, the set
\begin{equation}
\{\rho[U] | U\in U(n)\}   
\end{equation}
represents all partially coherent waves with the same total power and coherence spectrum as $\rho$. This set is entirely determined by $\bm{\lambda}^\downarrow(\rho)$.

Under unitary control, the absorptivity becomes explicitly $U$-dependent: 
\begin{align}
\label{eq:alpha_U}
\alpha \to \alpha [U] = \operatorname{tr} (U \rho U^\dagger  A). 
\end{align}
Unitary control of absorption for coherent waves has been studied in Ref.~\cite{guo2023b}. Here, we extend this method to manipulate the absorption of partially coherent waves.

\subsection{Major questions}

We ask two basic questions: Given a medium and a partially coherent incident wave, under unitary control, (1) What absorptivity is attainable? (2) How to obtain a given absorptivity? Question 1 inquires about the capability and limitations of unitary control. Question 2 seeks an implementation.

We now reformulate these key questions mathematically. Question 1: Given $S$ and $\rho$, what is the set
\begin{align}
 \{\alpha\} &\equiv \Set{\alpha[U] | U\in U(n)}?\label{eq:Question1-1} 
\end{align}
(If the density matrix $\rho$ needs to be specified, we denote $\alpha [U]$ as $\alpha [U|\rho]$ and $\{\alpha\}$ as $\{\alpha | \rho\}$.)

Question 2: Given $S$, $\rho$, and $\alpha_0 \in \{\alpha\}$, find a $U_0 \in U(n)$ such that
\begin{equation}
    \alpha[U_0] = \alpha_0.\label{eq:set_U_alpha0}
\end{equation}

\subsection{Main results}

In this subsection, we provide complete answers to Questions 1 and 2.

\subsubsection{Answer to Question 1}

We start with Question 1. The answer is
\begin{equation}\label{eq:main_result_set}
\{\alpha\} = \left[\bm{\lambda}^\downarrow(\rho)\cdot \bm{\lambda}^\uparrow(A), \bm{\lambda}^\downarrow(\rho)\cdot \bm{\lambda}^\downarrow(A) \right],     
\end{equation}
where $[\,,\,]$ denotes the closed real interval, and $\cdot$ denotes the usual inner product.
\begin{proof}
$\alpha[U]$ is a continuous map from $U(n)$ to $\mathbb{R}$. Since $U(n)$ is connected, the image of $\alpha[U]$, i.e., $\{\alpha\}$ is an interval. First, we show that for any $\alpha[U]\in \{\alpha\}$,
\begin{equation}\label{eq:proof1_goal}
\bm{\lambda}^\downarrow(\rho)\cdot \bm{\lambda}^\uparrow(A)
    \leq \alpha[U] \leq \bm{\lambda}^\downarrow(\rho)\cdot \bm{\lambda}^\downarrow(A).   
\end{equation} 
\red{We use the following theorem~\cite{richter1958,mirsky1959,ruhe1970,theobald1975}: If $M$ and $N$ are $n\times n$ Hermitian matrices, then
\begin{equation}\label{eq:theorem_trace}
\bm{\lambda}^{\downarrow}(M)\cdot \bm{\lambda}^{\uparrow}(N) \leq \operatorname{tr} MN \leq \bm{\lambda}^{\downarrow}(M)\cdot \bm{\lambda}^{\downarrow}(N).    
\end{equation}
Applying Eq.~(\ref{eq:theorem_trace}) to Eq.~(\ref{eq:alpha_U}), we obtain}
\begin{equation}\label{eq:proof_step1}
\bm{\lambda}^\downarrow(\rho)\cdot\bm{\lambda}^\uparrow(U^{\dagger}AU)
    \leq \alpha[U] \leq \bm{\lambda}^\downarrow(\rho)\cdot \bm{\lambda}^\downarrow(U^{\dagger}AU).    
\end{equation} 
Note that $\bm{\lambda}(A)$ is invariant under unitary control:
\begin{equation}\label{eq:proof_step2}
\bm{\lambda}(U^\dagger A U) = \bm{\lambda}(A).
\end{equation}
Substituting Eq.~(\ref{eq:proof_step2}) into Eq.~(\ref{eq:proof_step1}) yields Eq.~(\ref{eq:proof1_goal}).

Second, we show that the entire interval in Eq.~(\ref{eq:main_result_set}) is achievable by unitary control.  It suffices to demonstrate that both endpoints are achievable. Since $\rho$ and $A$ are Hermitian, they are diagonalizable by unitary similarity~\cite{horn2012}:
\begin{align}\label{eq:diagonalize_rho_A}
\rho = U_1 D_\rho^\downarrow U_1^\dagger,  \quad
A = V_1 D_A^\downarrow V_1^\dagger = V_2 D_A^\uparrow V_2^\dagger,
\end{align}
where $U_1$, $V_1$, $V_2 \in U(n)$, and 
\begin{align}
&D_\rho^\downarrow = \operatorname{diag}[\bm{\lambda}^\downarrow(\rho)], \\
&D_A^\downarrow = \operatorname{diag}[\bm{\lambda}^\downarrow(A)], \; 
D_A^\uparrow = \operatorname{diag}[\bm{\lambda}^\uparrow(A)],
\end{align}
\red{where $\operatorname{diag}(\bm{v})$ represents a diagonal matrix with the elements of vector $\bm{v}$ along its main diagonal.} We define the unitary matrices
\begin{equation}\label{eq:U_u_U_l}
U_u = V_1 U_1^\dagger, \quad U_l = V_2 U_1^\dagger.     
\end{equation}
We can then verify that
\begin{align}
\alpha[U_u] &= \operatorname{tr} (\rho U_u^\dagger  AU_u)
= \operatorname{tr} (U_1 D_\rho^\downarrow U_1^\dagger  U_1 V_1^\dagger V_1 D_A^\downarrow V_1^\dagger V_1 U_1^\dagger)  \notag \\ &= \operatorname{tr} (D_\rho^\downarrow D_A^\downarrow) = \bm{\lambda}^\downarrow(\rho)\cdot \bm{\lambda}^\downarrow(A).   
\end{align}
\begin{align}
\alpha[U_l] &= \operatorname{tr} (\rho U_l^\dagger  AU_l)
= \operatorname{tr} (U_1 D_\rho^\downarrow U_1^\dagger  U_1 V_2^\dagger V_2 D_A^\uparrow V_2^\dagger V_2 U_1^\dagger)  \notag \\ &= \operatorname{tr} (D_\rho^\downarrow D_A^\uparrow) = \bm{\lambda}^\downarrow(\rho)\cdot \bm{\lambda}^\uparrow(A).   
\end{align} 
Hence both endpoints are achievable by unitary control. This completes the proof of Eq.~(\ref{eq:main_result_set}).
\end{proof}
Eq.~(\ref{eq:main_result_set}) is the first main result of our paper. It completely characterizes the attainable absorptivity via unitary control. It shows that $\{\alpha\}$ is completely determined by $\bm{\lambda}(\rho)$ and $\bm{\lambda}(A)$, which are invariant under unitary control. 

\begin{figure}[hbtp]
    \centering
    \includegraphics[width=0.48\textwidth]{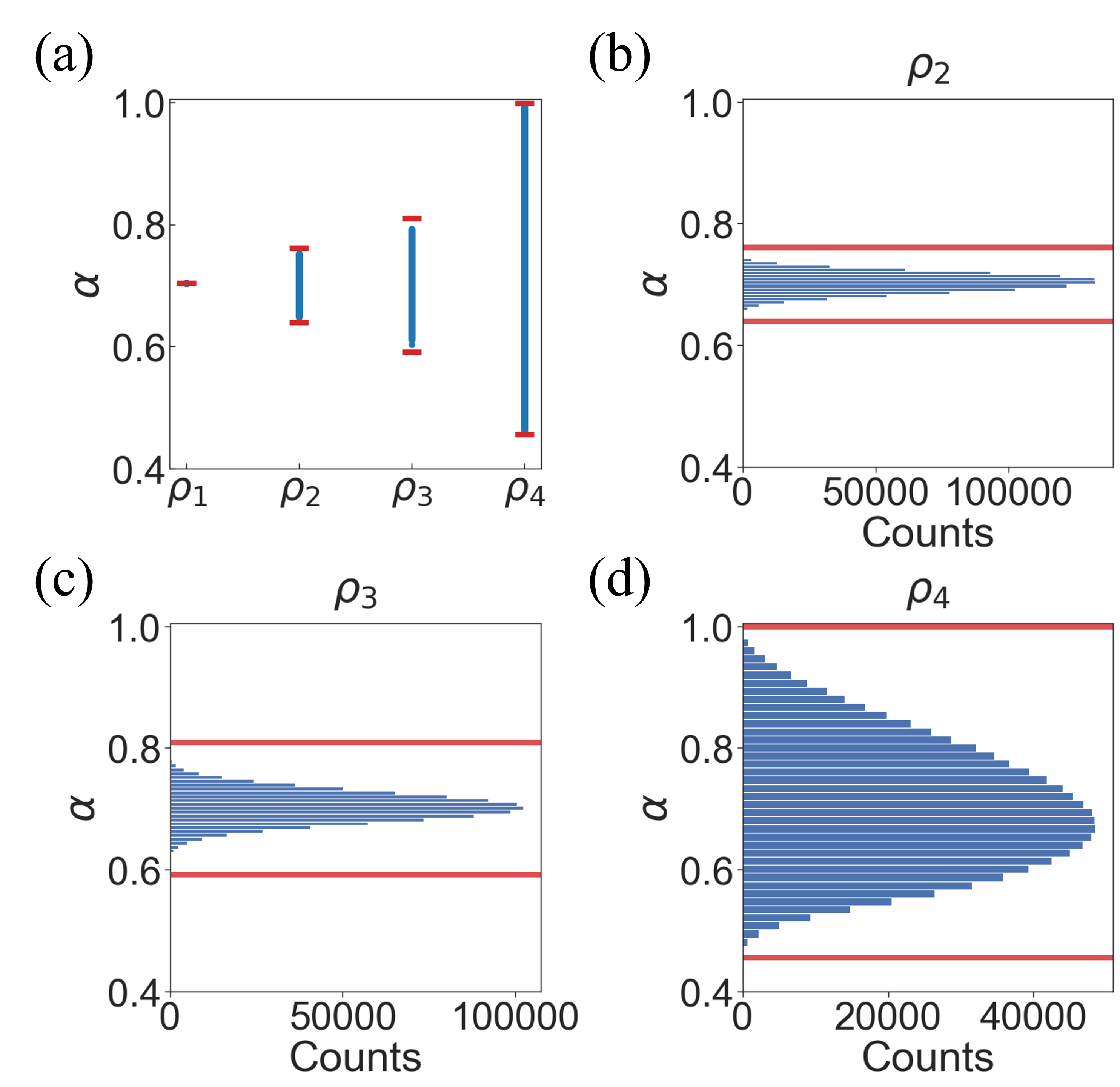}
    \caption{Attainable absorptivity under unitary control [Eq.~(\ref{eq:main_result_set})]. (a) Blue dots: $\alpha[U_i|\rho_j]$ for $1,000,000$ random unitary matrices $U_i$ and input density matrices $\rho_j$ with $j=1,2,3,4$. Red lines: calculated interval endpoints by Eq.~(\ref{eq:main_result_set}). (b-d) Histograms of $\alpha[U_i|\rho_j]$ for $j=2,3,4$. $\alpha[U_i|\rho_1]$ is constant for all $U_i$, as $\rho_1$ corresponds to an incoherent input wave. }
    \label{fig:numerical}
\end{figure}

To illustrate our results, we conduct a numerical experiment. \red{We generate a random $5\times 5$ scattering matrix using NumPy's random number generator~\cite{harris2020}:}
\begin{widetext}
\begin{equation}\label{eq:example1_S_matrix}
S = 
\begin{pmatrix}
0.03-0.18 i & 0.02 - 0.15 i & -0.01-0.09 i & -0.04 - 0.12 i & 0.14 - 0.01 i \\
0.32+0.02 i & 0.06-0.04 i & 0.15 - 0.02 i & -0.10 +0.00i & -0.40-0.01 i \\
0.01 - 0.24 i & -0.29 + 0.03 i & 0.11 - 0.07 i & 0.14 - 0.24 i & -0.40 - 0.03 i \\
-0.11 - 0.02 i & -0.07 - 0.02 i & -0.13 - 0.11 i & -0.37 - 0.34 i & -0.02 - 0.27 i \\
-0.16+0.11i & -0.08+0.34i & 0.01 + 0.18 i & 0.31 + 0.01 i & -0.13-0.07 i 
\end{pmatrix}.
\end{equation}    
\end{widetext}
\red{This matrix is non-unitary and has:} 
\begin{align}
\bm{\sigma}^\downarrow(S) &= (0.74,0.71,0.66,0.09,0.03) \\
\bm{\lambda}^\uparrow(A) &= 1 - \bm{\sigma}^{2\downarrow}(S) = (0.46, 0.50, 0.57, 0.99, 1.00). 
\end{align}
We consider four input density matrices $\rho_1$, $\rho_2$, $\rho_3$, and $\rho_4$, with coherence spectra:
\begin{align}
\bm{\lambda}^\downarrow(\rho_1) &= (0.20, 0.20, 0.20, 0.20, 0.20), \label{eq:lambda_rho_1}\\ \bm{\lambda}^\downarrow(\rho_2) &= (0.25, 0.25, 0.22, 0.18, 0.09), \label{eq:lambda_rho_2}\\
\bm{\lambda}^\downarrow(\rho_3) &= (0.34, 0.26, 0.21, 0.11, 0.08), \label{eq:lambda_rho_3}\\
\bm{\lambda}^\downarrow(\rho_4) &= (1.00, 0.00, 0.00, 0.00, 0.00). \label{eq:lambda_rho_4}
\end{align}
Note that $\rho_1$ is completely incoherent, $\rho_2$ and $\rho_3$ are partially coherent, and $\rho_4$ is perfectly coherent. For each input, we generate $1,000,000$ random unitary matrices $U_i$ from Circular Unitary Ensemble with Haar measure~\cite{mezzadri2007}, which provides a uniform probability distribution on $U(n)$~\cite{mezzadri2007}. Then we calculate the absorptivity $\alpha[U_i|\rho_j]= \operatorname{tr} (U_i \rho_j U_i^\dagger  A)$ for each $\rho_j$ by Eq.~(\ref{eq:alpha_U}). Fig.~\ref{fig:numerical}a shows that, all the $\alpha[U_i|\rho_j]$ are contained within the corresponding intervals $\{\alpha|\rho_j\}$ as determined by Eq.~(\ref{eq:main_result_set}). Moreover, the plot suggests that the entire interval can be covered when all unitary matrices in $U(n)$ are considered. 
We check that 
\begin{align}
\{\alpha|\rho_1\} &= \{\frac{1}{5} \sum_{i=1}^{5}\lambda_i(A)\} = \{0.704\},  \\
\{\alpha|\rho_2\} &= \left[0.64, 0.76\right],  \\
\{\alpha|\rho_3\} &= \left[0.59, 0.81\right], \\
\{\alpha|\rho_4\} &= \left[1-\sigma^2_1(S), 1-\sigma^2_n(S)\right] = \left[0.46,1.00\right]. 
\end{align}
Figs.~\ref{fig:numerical}b-\ref{fig:numerical}d show the histograms of $\alpha[U_i]$ for $\rho_2, \rho_3$, and $\rho_4$, respectively. We observe that the probability distribution of $\alpha[U_i]$ depends on $\bm{\lambda}(\rho)$. We also verify through additional simulations that $\bm{\sigma}(S)$ affects the probability distribution of $\alpha[U_i]$ as well. The explicit dependence of the probability distribution of $\alpha[U_i]$ on $\bm{\lambda}(\rho)$ and $\bm{\sigma}(S)$ is an interesting open question.

\subsubsection{Answer to Question 2}

Now we turn to Question~2. The problem corresponds to the following physical scenario.  Suppose we have a photonic structure characterized by a scattering matrix $S$ and, consequently, an absorptivity matrix $A = I - S^\dagger S$. Consider an incident partially coherent wave characterized by a normalized density matrix $\rho$. Given a absorptivity 
\begin{equation}\label{eq:alpha0_in_bound}
    \bm{\lambda}^\downarrow(\rho)\cdot \bm{\lambda}^\uparrow(A)
    \leq \alpha_0 \leq \bm{\lambda}^\downarrow(\rho)\cdot \bm{\lambda}^\downarrow(A),
\end{equation}
how do we construct one unitary control scheme described by a unitary matrix $U[\alpha_0]$ that achieves $\alpha_0$? 

We solve this problem using the following algorithm:
\begin{algorithm}[Constructing a ${U[\alpha_{0}]}$]\label{alg:U_alpha0}
\hfill
\begin{enumerate}
\item Diagonalize $\rho$ and $A$ as in Eq.~(\ref{eq:diagonalize_rho_A}). \item Construct $U_u$ and $U_l$ according to Eq.~(\ref{eq:U_u_U_l}). 
\item Define a skew-Hermitian matrix
\begin{equation}
    J = \log (U_u U_l^\dagger),
\end{equation}
and construct a continuous path between $U_l$ and $U_u$ in $U(n)$:
\begin{equation}
    U(\tau) \coloneqq e^{J \tau} U_l, \quad 0 \le \tau \le 1.
\end{equation} 
\red{The function $\tau \mapsto \alpha[U(\tau)]$ is continuous on the interval $[0,1]$, with $\alpha[U(0)] = \bm{\lambda}^\downarrow(\rho)\cdot \bm{\lambda}^\uparrow(A)$ and $\alpha[U(1)] = \bm{\lambda}^\downarrow(\rho)\cdot \bm{\lambda}^\downarrow(A)$.}
By continuity, the values of $\alpha[U(\tau)]$ for $0 \le \tau \le 1$ will cover the whole interval of $\{\alpha\}$. \item Use the bisection search to find $0 \le \tau_0 \le 1$ such that
\begin{equation}
    \alpha[U(\tau_0)] = \alpha_0.
\end{equation}
Thus we obtain a $U[\alpha_0] = U(\tau_0)$.
\end{enumerate}
\end{algorithm}

\begin{figure}
    \centering
    \includegraphics[width=0.35\textwidth]{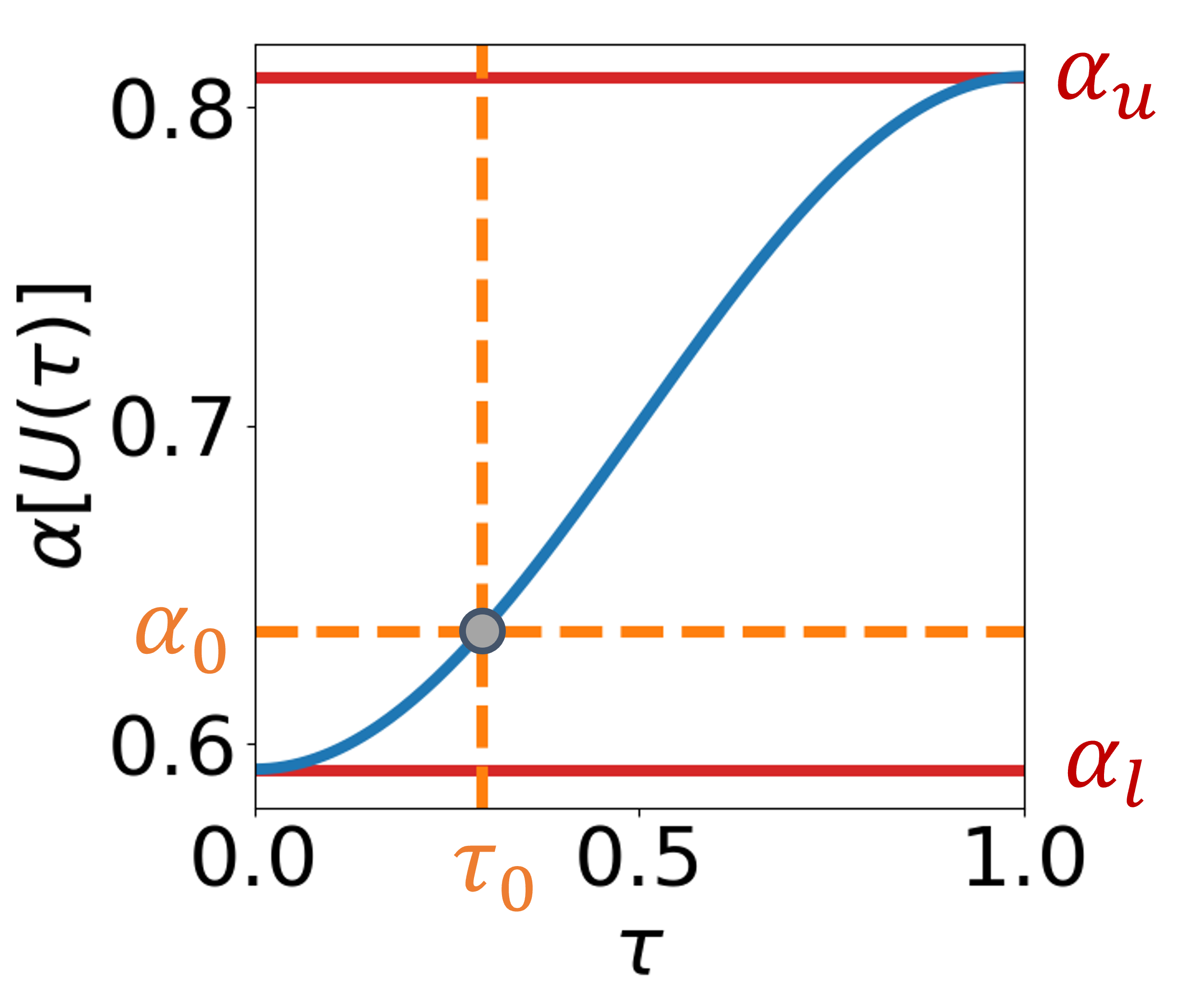}
    \caption{Constructing a unitary matrix for desired absorptivity [Algorithm~\ref{alg:U_alpha0}]. Blue curve: $\alpha[U(\tau)]$, $\tau$ is a parameter that varies from $0$ to $1$, corresponding to a continuous path between the unitary matrices $U_l$ and $U_u$. $U(\tau_0)$ achieves a desired absorptivity $\alpha_0 \in [\alpha_l, \alpha_u]$. }
    \label{fig:algorithm}
\end{figure}

Algorithm~\ref{alg:U_alpha0} is our second main result. To illustrate its usage, we provide a numerical example. We consider the same $S$-matrix in Eq.~(\ref{eq:example1_S_matrix}) and the input density matrix $\rho_3$ as introduced in Eq.~(\ref{eq:lambda_rho_3}). Our task is to construct a $U[\alpha_0]$ with a randomly assigned goal
\begin{equation}
     0.6355 = \alpha_0 \in \{\alpha|\rho_3\} = \left[0.59, 0.81\right].
\end{equation}
We use Algorithm~\ref{alg:U_alpha0} and obtain
\begin{widetext}
\begin{equation}\label{eq:U_opt}
U[\alpha_0] = 
\begin{pmatrix}
-0.26 + 0.31 i & -0.51 - 0.51 i & 0.08 -0.28 i & -0.16 + 0.18 i & -0.15 + 0.40 i \\
-0.25 -0.26 i & -0.23 + 0.06 i & -0.59 + 0.37 i & 0.02 + 0.27i & -0.48 - 0.09 i \\
0.51 + 0.03 i & -0.33 - 0.17 i & 0.41 + 0.43 i & -0.18 + 0.35 i & -0.07 - 0.36 i \\
-0.47 + 0.41 i & -0.04 + 0.45 i & 0.20 - 0.14 i & -0.39 + 0.02 i & -0.15 - 0.44 i \\
-0.30 - 0.15 i & 0.28 - 0.07 i & 0.06 + 0.01 i & 0.08 + 0.72 i & 0.48 -0.03 i 
\end{pmatrix}.
\end{equation}    
\end{widetext}

\section{Applications}\label{sec:applications}

Now we discuss the physical applications of our theory. 

\subsection{Partially coherent perfect absorption}

First, we introduce the phenomenon of \emph{partially coherent perfect absorption}. Recall that coherent perfect absorption refers to the effect that a coherent wave is perfectly absorbed by a linear system via unitary control~\cite{chong2010a,guo2023b}. For a linear system characterized by a scattering matrix $S$, coherent perfect absorption occurs if and only if
\begin{equation}\label{eq:criterion_CPA}
\operatorname{nullity} S \ge 1.  
\end{equation}
Similarly, partially coherent perfect absorption refers to the phenomenon where a partially coherent wave is perfectly absorbed by a linear system via unitary control. We apply our theory to prove the following criterion: For a linear system characterized by a scattering matrix $S$ and a partially coherent wave characterized by a density matrix $\rho$, partially coherent perfect absorption occurs if and only if
\begin{equation}\label{eq:null_rank}
\operatorname{nullity} S \ge \operatorname{rank} \rho.  
\end{equation}
As a sanity check, for a perfectly coherent wave, $\operatorname{rank}\rho =1$, and Eq.~(\ref{eq:null_rank}) reduces to Eq.~(\ref{eq:criterion_CPA}).  

\begin{proof}
Note that $\operatorname{nullity} S$ equals the number of zeros in $\bm{\sigma}(S)$, thus the number of ones in $\bm{\lambda}(A)$ [see Eq.~(\ref{eq:lamda_A_sigma_S})], while $\operatorname{rank} \rho$ equals the number of non-zero elements in $\bm{\lambda}(\rho)$. Hence,
\begin{align}\label{eq:lambda_A_nullity}
\bm{\lambda}^\downarrow(A) &= (\overbrace{1,\ldots,1}^{\operatorname{nullity} S}, \overbrace{*, \ldots, *}^{0\leq *<1}), \\
\bm{\lambda}^\downarrow(\rho) &= (\underbrace{*,\ldots,*}_{0<*\leq 1}, \underbrace{0, \ldots, 0}_{1-\operatorname{rank}\rho}).\label{eq:lambda_rho_rank} 
\end{align}
According to the normalization condition [Eq.~(\ref{eq:rho_normalization})], 
\begin{equation}\label{eq:lambda_rho_normalization}
\sum_i \lambda^\downarrow_i(\rho) = \operatorname{tr}\rho = 1.    
\end{equation}
By Eq.~(\ref{eq:main_result_set}), the maximal absorptivity achievable via unitary control is 
\begin{equation}
\alpha_{\max} = \bm{\lambda}^\downarrow(\rho)\cdot \bm{\lambda}^\downarrow(A). 
\end{equation}
Partially coherent perfect absorption occurs iff $\alpha_{\max} =1$. Using Eqs.~(\ref{eq:lambda_A_nullity}), (\ref{eq:lambda_rho_rank}), and (\ref{eq:lambda_rho_normalization}), we obtain that 
\begin{equation}
\alpha_{\max} =1 \iff \operatorname{nullity} S \ge \operatorname{rank} \rho.      
\end{equation}
This completes the proof of the criterion (\ref{eq:null_rank}). 
\end{proof}

If the criterion (\ref{eq:null_rank}) is satisfied, we can unitarily transform the input density matrix $\rho$ such that its support~\footnote{The support of a density matrix $\rho$ is the orthogonal complement of the kernel of $\rho$~\cite{robert2005}.} is a subset of the null space of the $S$-matrix, thus achieving partially coherent perfect absorption. $U_u$ as defined in Eq.~(\ref{eq:U_u_U_l}) provides one such unitary transformation. 

\begin{figure}[hbtp]
    \centering
    \includegraphics[width=0.5\textwidth]{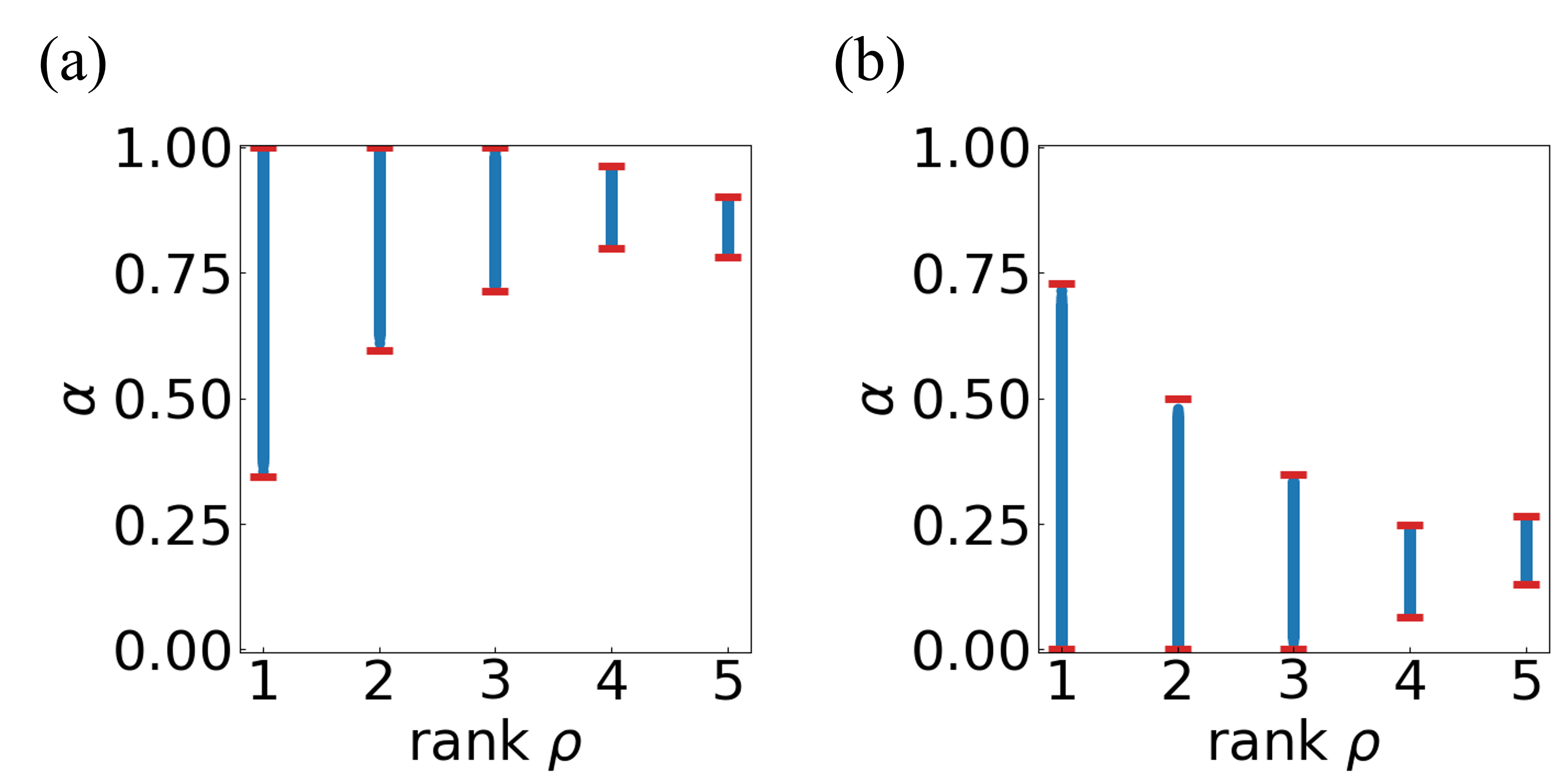}
    \caption{Numerical demonstration of the conditions for (a) partially coherent perfect absorption and (b) partially coherent zero absorption. Blue dots: $\alpha[U_i|\tilde{\rho}_j]$ for $1,000,000$ random $U_i$ and $j=1,2,3,4,5$. Red lines: calculated interval endpoints by Eq.~(\ref{eq:main_result_set}).}
    \label{fig:perfect}
\end{figure}

We numerically demonstrate our results on partially coherent perfect absorption. We consider a $5\times 5$ lossy scattering matrix:
\begin{widetext}
\begin{equation}\label{eq:example2_S_matrix}
{S} = 
\begin{pmatrix}
-0.16 + 0.01 i & 0.09 - 0.02 i & -0.04 - 0.08 i & -0.05 + 0.00 i & -0.07 + 0.13 i \\
0.10+0.10 i & -0.08-0.17i & -0.16-0.16i & -0.20+0.00i &  0.08+0.07i\\
0.11+0.09i & -0.15+0.01i & -0.10+0.23i &  0.11+0.17i &  0.21-0.05i \\
-0.01+0.27i &  0.15-0.24i &  0.11-0.26i & -0.10-0.26i &  0.05+0.12i  \\
-0.03+0.14i &  0.06-0.10i &  0.02-0.08i & -0.02-0.07i &  0.04+0.08i
\end{pmatrix},
\end{equation}
\end{widetext}
which has 
\begin{align}
\bm{\sigma}^\downarrow({S}) &= (0.81,0.38,0.00,0.00,0.00), \\
\bm{\lambda}^\uparrow({A}) &= 1 - \bm{\sigma}^{2\downarrow}({S}) = (0.34, 0.85, 1.00, 1.00, 1.00),  
\end{align}
thus
\begin{equation}
\operatorname{nullity} {S} = 3. 
\end{equation}
We consider five different incident waves characterized by normalized density matrices $\tilde{\rho}_j$, $1 \le j \le 5$, with coherence spectra:
\begin{align}\label{eq:coherence_spectrum_rho_1}
\bm{\lambda}^\downarrow(\tilde{\rho}_1) &= (1.00, 0.00, 0.00, 0.00, 0.00), \\ \bm{\lambda}^\downarrow(\tilde{\rho}_2) &= (0.51, 0.49, 0.00, 0.00, 0.00), \\
\bm{\lambda}^\downarrow(\tilde{\rho}_3) &= (0.36, 0.33, 0.31, 0.00, 0.00), \\
\bm{\lambda}^\downarrow(\tilde{\rho}_4) &= (0.25, 0.25, 0.25, 0.25, 0.00), \\
\bm{\lambda}^\downarrow(\tilde{\rho}_5) &= (0.28, 0.23, 0.21, 0.16, 0.12),\label{eq:coherence_spectrum_rho_5}
\end{align}
thus their ranks are different:
\begin{equation}
\operatorname{rank} \tilde{\rho}_j = j.
\end{equation}
For each input, we generate $1,000,000$ random unitary matrices $U_i$ from the Circular Unitary Ensemble. Then we calculate the absorptivity $\alpha[U_i|\tilde{\rho}_j]$ for each $\tilde{\rho}_j$ by Eq.~(\ref{eq:alpha_U}) and plot the results in  Fig.~\ref{fig:perfect}a. We see that partially coherent perfect absorption is achievable when $\operatorname{rank} \tilde{\rho}_j = 1, 2, 3$, but is not when $\operatorname{rank} \tilde{\rho}_j = 4, 5$. This checks the criterion~(\ref{eq:null_rank}). 

\subsection{Partially coherent zero absorption}

Second, we introduce the phenomenon of \emph{partially coherent zero absorption}. Coherent zero absorption refers to the effect that a coherent wave exhibits zero absorption by a linear system via unitary control. For a linear system characterized by a scattering matrix $S$ and thus an absorptivity matrix $A = I -S^\dagger S$, coherent zero absorption occurs if and only if
\begin{equation}\label{eq:criterion_CZA}
\operatorname{nullity} A \ge 1.  
\end{equation}
Similarly, partially coherent zero absorption refers to the phenomenon where a partially coherent wave exhibits zero absorption by a linear system via unitary control. We apply our theory to prove the following criterion: For a linear system characterized by a scattering matrix $S$ and an absorptivity matrix $A = I -S^\dagger S$ and a partially coherent wave characterized by a density matrix $\rho$, partially coherent zero absorption occurs if and only if
\begin{equation}\label{eq:null_rank_CZA}
\operatorname{nullity} A \ge \operatorname{rank} \rho.  
\end{equation}
As a sanity check, for a perfectly coherent wave, $\operatorname{rank}\rho =1$, and Eq.~(\ref{eq:null_rank_CZA}) reduces to Eq.~(\ref{eq:criterion_CZA}).

\begin{proof}
Note that $\operatorname{nullity} A$ equals the number of zeros in $\bm{\sigma}(A)$, while $\operatorname{rank} \rho$ equals the number of non-zero elements in $\bm{\lambda}(\rho)$. Hence,
\begin{align}\label{eq:lambda_A_nullity_CZA}
\bm{\lambda}^\uparrow(A) &= (\overbrace{0,\ldots,0}^{\operatorname{nullity} A}, \overbrace{*, \ldots, *}^{0< *\leq1}), \\
\bm{\lambda}^\downarrow(\rho) &= (\underbrace{*,\ldots,*}_{0<*\leq 1}, \underbrace{0, \ldots, 0}_{1-\operatorname{rank}\rho}).\label{eq:lambda_rho_rank_CZA} 
\end{align}
By Eq.~(\ref{eq:main_result_set}), the minimal absorptivity achievable via unitary control is 
\begin{equation}
\alpha_{\min} = \bm{\lambda}^\downarrow(\rho)\cdot \bm{\lambda}^\uparrow(A). 
\end{equation}
Partially coherent zero absorption occurs iff $\alpha_{\min} =0$. Using Eqs.~(\ref{eq:lambda_A_nullity_CZA}), and (\ref{eq:lambda_rho_rank_CZA}), we obtain that 
\begin{equation}
\alpha_{\min} = 0 \iff \operatorname{nullity} A \ge \operatorname{rank} \rho.      
\end{equation}
This completes the proof of the criterion (\ref{eq:null_rank_CZA}). 
\end{proof}

If the criterion (\ref{eq:null_rank_CZA}) is satisfied, we can unitarily transform the input density matrix $\rho$ such that its support is a subset of the null space of the $A$-matrix, thus achieving partially coherent zero absorption. $U_l$ as defined in Eq.~(\ref{eq:U_u_U_l}) provides one such unitary transformation.

We numerically demonstrate our results on partially coherent zero absorption. We consider a $5\times 5$ lossy scattering matrix:
\begin{widetext}
\begin{equation}\label{eq:example3_S_matrix}
S = 
\begin{pmatrix}
0.17 - 0.09 i & -0.44 - 0.34 i & -0.33 - 0.10 i & 0.16 - 0.46 i & 0.24 - 0.27 i \\
0.06 -0.05 i & 0.55 - 0.12i & -0.07 + 0.68i & 0.12 - 0.20i &  0.12 + 0.07i\\
-0.11 + 0.25i & 0.29 - 0.23i & -0.36 - 0.12i &  -0.63 - 0.13i &  0.21 + 0.05i \\
0.66 - 0.13i &  -0.06 + 0.33i &  0.14 + 0.05i & -0.17 - 0.01i &  0.55 + 0.15i  \\
-0.28 + 0.31i &  -0.01 + 0.22i &  0.25 + 0.10i & -0.17 - 0.27i &  0.27 - 0.31i
\end{pmatrix},
\end{equation}
\end{widetext}
which has 
\begin{align}
\bm{\sigma}^\downarrow(S) &= (1.00,1.00,1.00,0.86,0.52), \\
\bm{\lambda}^\uparrow(A) &= 1 - \bm{\sigma}^{2\downarrow}(S) = (0.00, 0.00, 0.00, 0.26, 0.73),  
\end{align}
thus
\begin{equation}
\operatorname{nullity} A = 3. 
\end{equation}
We consider five different incident waves characterized by normalized density matrices $\tilde{\rho}_j$, $1 \le j \le 5$, with coherence spectra as provided in Eqs.~(\ref{eq:coherence_spectrum_rho_1}-\ref{eq:coherence_spectrum_rho_5}); 
thus, $\operatorname{rank} \tilde{\rho}_j = j$.
For each input, we generate $1,000,000$ random unitary matrices $U_i$ from Circular Unitary Ensemble. Then we calculate the absorptivity $\alpha[U_i|\tilde{\rho}_j]$ for each $\tilde{\rho}_j$ by Eq.~(\ref{eq:alpha_U}) and plot the results in  Fig.~\ref{fig:perfect}b. We see that partially coherent zero absorption is achievable when $\operatorname{rank} \tilde{\rho}_j = 1, 2, 3$, but is not when $\operatorname{rank} \tilde{\rho}_j = 4, 5$. This checks the criterion~(\ref{eq:null_rank_CZA}). 

\subsection{Majorized coherence implies nested absorption intervals}

Third, we examine how the degree of coherence affects the attainable absorption interval. Our main result Eq.~(\ref{eq:main_result_set}) shows that, for a given lossy system, the absorptivity interval $\{\alpha\}$ is solely controlled by the coherence spectrum $\bm{\lambda}^{\downarrow}(\rho)$. A natural question arises: How will the absorptivity interval vary when the degree of coherence changes?

To address this question, we must first clarify how to compare the coherence between waves. A natural measure is provided by majorization~\cite{marshall2011,nielsen1999,gour2015,bengtsson2017,gour2018,luis2016}: For two vectors $\bm{x}$ and $\bm{y}$ in $\mathbb{R}^n$, we say that $\bm{x}$ is majorized by $\bm{y}$, written as $\bm{x} \prec \bm{y}$, if
\begin{align}
&\sum_{i=1}^k x_i^\downarrow \le \sum_{i=1}^k y_i^\downarrow, \quad k=1,2,\ldots,n-1;     \\
&\sum_{i=1}^n x_i = \sum_{i=1}^n y_i.\label{eq:def_majorization_equality}
\end{align}
Intuitively, $\bm{x} \prec \bm{y}$ means that their components have the same sum, but the components of $\bm{x}$ are no more spread out than those of $\bm{y}$. Consider two waves with density matrices $\rho_1$ and $\rho_2$, respectively. We say that $\rho_1$ is no more coherent than $\rho_2$ if $\bm{\lambda}^{\downarrow}(\rho_1) \prec \bm{\lambda}^{\downarrow}(\rho_2)$. If neither $\bm{\lambda}^{\downarrow}(\rho_1) \prec \bm{\lambda}^{\downarrow}(\rho_2)$ nor $\bm{\lambda}^{\downarrow}(\rho_2) \prec \bm{\lambda}^{\downarrow}(\rho_1)$ holds, we say that $\rho_1$ and $\rho_2$ are incomparable and write $\bm{\lambda}^{\downarrow}(\rho_1) \parallel \bm{\lambda}^{\downarrow}(\rho_2)$. As a sanity check, for any $\rho$,
\begin{equation}\label{eq:incoherent_partial_coherent}
(\frac{1}{n},\frac{1}{n},\ldots,\frac{1}{n}) \prec \bm{\lambda}^\downarrow(\rho) \prec (1,0,\ldots,0),
\end{equation}
i.e., any wave is no more coherent than a coherent wave, and no less coherent than an incoherent wave.

Now we can state the following theorem: If $\rho_1$ is no more coherent than $\rho_2$, then for any lossy system, the absorption interval of $\rho_1$ is always contained in that of $\rho_2$:
\begin{equation}\label{eq:nested_interval}
\bm{\lambda}^\downarrow(\rho_1) \prec \bm{\lambda}^\downarrow(\rho_2) \implies \{\alpha\}_1 \subseteq \{\alpha\}_2.   
\end{equation}
Using Eq.~(\ref{eq:main_result_set}), we can express the right hand side of Eq.~(\ref{eq:nested_interval}) more explicitly:
\begin{equation}\label{eq:nested_interval_explicit}
\bm{\lambda}^\downarrow(\rho_2)\cdot \bm{\lambda}^\uparrow(A)\leq \bm{\lambda}^\downarrow(\rho_1)\cdot \bm{\lambda}^\uparrow(A) \leq 
\bm{\lambda}^\downarrow(\rho_1)\cdot \bm{\lambda}^\downarrow(A) \leq \bm{\lambda}^\downarrow(\rho_2)\cdot \bm{\lambda}^\downarrow(A)
\end{equation}
\begin{proof}
\red{We use the following theorem: Let $\mathcal{D} = \{(x_1,\ldots,x_n) \in \mathbb{R}^n: x_1 \ge \ldots \ge x_n\}$. The inequality
\begin{equation}
x \cdot u \leq  y \cdot u 
\end{equation}
holds for all $u\in \mathcal{D}$ if and only if $x \prec y$ on $\mathcal{D}$. (See Ref.~\cite{marshall2011}, p.~160, Proposition B.7.)} 

First, since $\bm{\lambda}^\downarrow(\rho_1)$, $\bm{\lambda}^\downarrow(\rho_2)$, and  $\bm{\lambda}^\downarrow(A)$ are all in $\mathcal{D}$, the theorem above implies that
\begin{equation}
\bm{\lambda}^\downarrow(\rho_1)\cdot \bm{\lambda}^\downarrow(A) \leq \bm{\lambda}^\downarrow(\rho_2)\cdot \bm{\lambda}^\downarrow(A),
\end{equation}
which proves the last inequality in (\ref{eq:nested_interval_explicit}). Second, since $-\bm{\lambda}^\uparrow(A) \in \mathcal{D}$, the theorem above implies that 
\begin{equation}
-\bm{\lambda}^\downarrow(\rho_1)\cdot \bm{\lambda}^\uparrow(A) \leq -\bm{\lambda}^\downarrow(\rho_2)\cdot \bm{\lambda}^\uparrow(A), 
\end{equation}
which is equivalent to the first inequality in (\ref{eq:nested_interval_explicit}). The middle inequality in (\ref{eq:nested_interval_explicit}) always holds. This completes the proof.
\end{proof}
Theorem (\ref{eq:nested_interval}) is our main result of this subsection. It can be summarized as: ``Majorized coherence implies nested absorption intervals." Now we examine its implications. 

First, we apply Theorem (\ref{eq:nested_interval}) to (\ref{eq:incoherent_partial_coherent}) and deduce that for any density matrix $\rho$ and any absorptivity matrix $A$,
\begin{equation}\label{eq:nested_interval_explicit_coherent}
\lambda_{\min}(A) \leq \bm{\lambda}^\downarrow(\rho)\cdot \bm{\lambda}^\uparrow(A) \leq  \frac{1}{n}\sum_i \lambda_i(A) \leq 
\bm{\lambda}^\downarrow(\rho)\cdot \bm{\lambda}^\downarrow(A) \leq \lambda_{\max}(A). 
\end{equation}
In particular, the mean of $\lambda_i(A)$ is always contained in the absorption interval and thus attainable via unitary control.  

Second, from the contrapositive of Theorem (\ref{eq:nested_interval}), we deduce that if for some lossy system, neither $\{\alpha\}_1 \subseteq \{\alpha\}_2$ nor $\{\alpha\}_2 \subseteq \{\alpha\}_1$ holds (denoted as $\{\alpha\}_1 \parallel \{\alpha\}_2$), then $\rho_1$ and $\rho_2$ are incomparable:
\begin{equation}\label{eq:incomparable_interval}
\{\alpha\}_1 \parallel \{\alpha\}_2 \implies   \bm{\lambda}^\downarrow(\rho_1) \parallel \bm{\lambda}^\downarrow(\rho_2). 
\end{equation}

We illustrate these results with previous numerical examples. In Fig.~\ref{fig:numerical}a, we observe that
\begin{equation}
 \{\alpha|\rho_1\} \subseteq   \{\alpha|\rho_2\} \subseteq
 \{\alpha|\rho_3\} \subseteq
   \{\alpha|\rho_4\},    
\end{equation} 
because $\bm{\lambda}^\downarrow(\rho_i)$ as given in Eqs.~(\ref{eq:lambda_rho_1})-(\ref{eq:lambda_rho_4}) satisfy
\begin{equation}
\bm{\lambda}^\downarrow(\rho_1) \prec \bm{\lambda}^\downarrow(\rho_2) \prec \bm{\lambda}^\downarrow(\rho_3) \prec \bm{\lambda}^\downarrow(\rho_4).
\end{equation}
In Fig.~\ref{fig:perfect}a and \ref{fig:perfect}b, we observe that
\begin{equation}
 \{\alpha|\tilde{\rho}_1\} \subseteq   \{\alpha|\tilde{\rho}_2\} \subseteq
 \{\alpha|\tilde{\rho}_3\} \subseteq
   \{\alpha|\tilde{\rho}_4\},    
\end{equation} 
because $\bm{\lambda}^\downarrow(\tilde{\rho}_i)$ as given in Eqs.~(\ref{eq:coherence_spectrum_rho_1})-(\ref{eq:coherence_spectrum_rho_5}) satisfy
\begin{equation}
\bm{\lambda}^\downarrow(\tilde{\rho}_1) \prec \bm{\lambda}^\downarrow(\tilde{\rho}_2) \prec \bm{\lambda}^\downarrow(\tilde{\rho}_3) \prec \bm{\lambda}^\downarrow(\tilde{\rho}_4).
\end{equation}
We also observe that
\begin{equation}
 \{\alpha|\tilde{\rho}_4\} \parallel  \{\alpha|\tilde{\rho}_5\},
\end{equation}
which can occur because 
\begin{equation}
\bm{\lambda}^\downarrow(\tilde{\rho}_4) \parallel \bm{\lambda}^\downarrow(\tilde{\rho}_5).
\end{equation}

\section{Conclusion}\label{sec:conclusion}

In conclusion, we have developed a systematic theory for unitary control of partially coherent wave absorption by linear systems. Our key results include (1) an analytical expression [Eq.~(\ref{eq:main_result_set})] for the range of attainable absorptivity under arbitrary unitary control of the incident field; and (2) an explicit algorithm (Algorithm~\ref{alg:U_alpha0}) to construct a unitary control scheme that achieves any desired absorptivity within the attainable range.

Through this theory, we obtain both fundamental insights and practical prescriptions. Fundamentally, we establish two new absorption phenomena, partially coherent perfect absorption and partially coherent zero absorption, and derive precise criteria [Eqs.~(\ref{eq:null_rank}) and (\ref{eq:null_rank_CZA})] for their occurrence. We also prove a fundamental theorem [Eq.~(\ref{eq:nested_interval})] relating the majorization order of the incident coherence spectra to the nesting order of the resulting absorption intervals. Practically, our algorithm provides step-by-step instructions to realize unitary control and obtain any physically allowed absorption outcome for a given incident partially coherent field and absorbing system.

The theory established in this work deepens the understanding of partially coherent absorption control and provides a powerful, flexible framework to engineer the absorption of partially coherernt waves across a diverse range of wave systems. We anticipate that our results will find applications in areas such as energy harvesting, thermal emission control, imaging, and sensing, where partially coherent radiative transfer plays a central role.

\begin{acknowledgments}
This work is funded by the U.~S.~Department of Energy (Grant No.~DE-FG02-07ER46426), and by a Simons Investigator in Physics grant from the Simons Foundation. (Grant No.~827065).

\end{acknowledgments}



\bibliography{main}

\end{document}